\documentclass[11pt]{article}
\usepackage{amssymb,amsfonts,amsmath,enumerate}
\usepackage{graphicx}
  \usepackage{amsmath}
    \usepackage{makeidx}
    \usepackage{amsfonts}
    \usepackage[ansinew]{inputenc}
    \usepackage[usenames,dvipsnames]{pstricks}
    \usepackage{subfigure}
    \usepackage{epsfig}
    \usepackage{pst-grad} 
    \usepackage{pst-plot} 
    \usepackage[colorlinks,hyperindex]{hyperref}
\usepackage{graphicx}
\usepackage{amssymb}
\usepackage{amsfonts}
\usepackage{graphicx}
\begin{document}

\title{Pattern Recognition In Non-Kolmogorovian Structures
}


\author{Federico Holik, Giuseppe Sergioli, Hector Freytes and Angelo Plastino }

\maketitle

\begin{abstract}
\noindent We present a generalization of the problem of pattern
recognition to arbitrary probabilistic models. This version deals
with the problem of recognizing an individual pattern among a family
of different species or classes of objects which obey probabilistic
laws which do not comply with Kolmogorov's axioms. We show that such
a scenario accommodates  many important examples, and in particular,
we provide a rigorous definition of the classical and the quantum
pattern recognition problems, respectively. Our framework allows for
the introduction of non-trivial correlations (as entanglement or
discord) between the different species involved, opening the door to
a new way of harnessing these physical resources for solving pattern
recognition problems. Finally, we present some examples and discuss
the computational complexity of the quantum pattern recognition
problem, showing that the most important quantum computation
algorithms can be described as non-Kolmogorovian pattern recognition
problems.\\
Quantum Pattern Recognition - Quantum Algorithms - Convex
Operational Models
\end{abstract}

\section{Introduction}\label{s:Introduction}

\noindent Pattern recognition is an active field of research which
has many applications in different disciplines, such as information
science, economics, engineering, and machine learning
\cite{Bishop-Book,TaiLiang-2016}. Intuitively, pattern recognition
could be defined as the problem of how a rational agent (which could
be an automata), decides to which class of objects a given new
object belongs. In its simpler version, given a family of classes of
objects $C_{i}$ (representing objects of different kinds), the
rational agent must decide to which class a given object $a$
belongs. The comparison is made with regards to a given set of
properties $\alpha_{j}$ of $a$. More sophisticated versions assume
that the knowledge of the rational agent is given in terms of
probability distributions. In its classical version, the properties
involved are compatible (can have definite simultaneous values) and
probabilities are Kolmogorovian (in the sense that they can be
described by the well known Kolmogorov's axioms
\cite{KolmogorovProbability}).

But it may happen that, for some particular models, the properties
involved cannot be determined simultaneously. This could be the case
if, for example, there is a limitation in the capability of
acquisition of knowledge by the agent. This could be originated in
epistemic constraints (for example, in game theory), or in
ontological limitations (as in quantum mechanics). Non-classical
effects have also been observed in cognitive phenomena (see for
example \cite{Aerts-2015,Aerts-2015a}). More concretely, suppose
that our rational agent deals with probabilistic models which do not
obey the laws of classical physics, in the sense that it is not
possible to attribute simultaneous values to the properties of the
objects involved \cite{Doring-KSVNA,Kochen-Specker}. How can we
formulate the pattern recognition problem in this non-Boolean
\cite{Adilee-CopulasQL,Svozil-2009} framework? The theory of pattern
recognition for this case cannot be the same as before, essentially,
due to the non-commutative character of the properties and the
probabilities involved. In other words, complementarity poses a
problem for the treatment of objects as possessing simultaneous
collections of well defined properties. For example, in the quantum
case, we must acknowledge the fact that the best way of describing a
class of objects is by attributing probabilities governed by the
laws of quantum mechanics
\cite{Gudder-StatisticalMethods,mikloredeilibro}. Then, observable
quantities will be represented mathematically by (possibly)
non-commutative operators acting on a Hilbert space \cite{vN}, and
this gives a different formulation of the discrimination problem
(for concrete examples of this, see
\cite{Sentis-CoherentStates-2015} and \cite{QPR3-Guta-Wojciech-NJP};
see also \cite{Moras-Sentis-Wittek-2016} for more discussion).

In order to describe how things work when probabilities depart from
the Kolmogorovian case, we present a formal quantum patterns
recognition's framework for generalized probabilistic models
\cite{Barnum-Wilce-2006,Barnum-Wilce-2010,Barnum-Wilce-2009}. In
this way we considerably expand the domain of applicability of this
field of research. Our main aim is to focus attention on the fact
that there are several versions of the problem, depending on the
structural aspects of the probabilities involved. Our theoretical
framework allows for introducing rigorous definitions of the
classical and quantum pattern recognition problems. At the same
time, it allows one to envisage the existence of other versions of
the problem, as it would appear, for example, in the relativistic or
thermodynamic limits.

It is important to remark that there are other approaches that use
non-classical techniques or quantum systems (like quantum computers)
to solve pattern recognition problems (see for example
\cite{QPR1-Horn-Gottlieb,Schuld-2015,Schutzhold-2003,QPR3-Trugenberger,QPR1}).
But they differ from our approach, mainly because the entities to be
discerned are classical (i.e., they do not exhibit quantum phenomena
such as superposition or entanglement). There also exist previous
formulations of the problem which are similar to ours for the
particular case of non-relativistic quantum mechanics and quantum
optics (see for example
\cite{Aimeur-QuantumWorld,QPR3-Guta-Wojciech-NJP,Moras-Sentis-Wittek-2016,QPR2-Sasaki-Carlini,QPR3-Sasaki-Carlini,Sentis-CoherentStates-2015}),
that can be naturally accommodated into our more general framework.
Our generalization could be useful for a better understanding of
these models, and, at the same time, it could serve as a suitable
tool for describing more general physical situations.

This perspective opens a field of research which is richer (from a
physical point of view) than its previous classical versions, mainly
because we allow for the possibility that the different classes of
objects involved display non-classical features, such as
complementarity, or non-classical correlations (such as entanglement
or discord). Things may change in a subtle manner when these
non-classical features cannot be neglected. This behavior would be
expected in any situation in which the elements involved in the
analysis are ``small" enough and reach the molecular or atomic
level. Of course, our approach can also be useful for classical
systems which are structured in such a way that, for one reason or
another, exhibit features that imitate quantum phenomena (this is
the case in some examples of game theory). Remarkably enough, a
connection between the study of relational databases and the
violation of Bell's inequalities is presented in \cite{Abramsky}.
This study suggests that some mathematical structures underlying
quantum contextuality can be found in fields of research in which
the data we might be dealing with is not necessarily about physical
objects. In this way, future developments of our mathematical
framework could be of use for the study of problems outside of
physics, such as relational databases and Big Data.

The paper is organized as follows. In Section
\ref{s:IntroPatternRcognition} we review the standard approach to
the pattern recognition problem. Next, in Section \ref{s:Generalized
Probabilities}, we review the different mathematical frameworks
which allow us to represent probabilities and properties which
depart from the Boolean-Commutative case. In Section
\ref{s:QuantumPatternRecognition} we present our version of the
pattern recognition problem for generalized probabilistic models,
and show how non-classical correlations can appear in the
non-Kolmogorovian probabilistic setting. In Section \ref{s:Examples}
we describe the particular cases of the standard, relativistic and
statistical quantum mechanical settings as concrete examples. We
study possible connections between quantum pattern recognition
theory and some relevant quantum algorithms in Section
\ref{s:Computational complexity}. Finally, in Section
\ref{s:Conclusions} we draw some conclusions.

\section{Classical Pattern Recognition Problem}\label{s:IntroPatternRcognition}

Suppose that, given a family of different classes $C_{i}$, a
rational agent (it could be a person or an automata) must decide to
which class a given individual $X$ pertains. It is important to
remark that the classes could be disjoint or not. For example, if
$C_{1}$ is a collection of dogs and $C_{2}$ is a collection of cats,
the aim of the agent is to decide, given an unknown individual $X$,
if it is a cat or a dog (from now on, following the jargon commonly used in pattern recognition, we use the terms ``object" and ``individual" interchangeably). It is usually assumed that knowledge about
the different classes and the given individual is given in terms of
a particular collection of properties (also called \emph{features})
of all possible individuals in question. The collection of
properties of a given individual $X^{i}_{j}\in C_{i}$ (individual
$j$ belonging to class $C_{i}$) is represented by an $n$-vector
$\vec{\alpha}^{i}_{j}=(\alpha^{i;j}_{1},\alpha^{i;j}_{2},...,\alpha^{i;j}_{n})$
(where the $\alpha^{i;j}_{k}$ take real values) and probability
distributions representing degrees of belief of the agent regarding
each individual having a particular collection of properties. This
is the most elementary form of the problem of \emph{pattern
recognition} \cite{Bishop-Book}. We will refer to it as the
\emph{classical pattern recognition problem}.

To be more specific, suppose that, given an individual $X$ to be
recognized, knowledge about it is represented by a probability
distribution $p(\vec{\alpha})$. Suppose also that knowledge about
each class $C_{i}$ is represented by a probability distribution
$p_{i}(\vec{\alpha}^{i}_{j})$ assigning a weight to each property
vector $\vec{\alpha}^{i}_{j}$ in $C_{i}$. Thus, the
\emph{classification problem}, is the problem of determining to
which class the individual is assigned to by contrasting knowledge
about the individual and the different classes. This can be done,
for example, by comparing the mean values of the considered
properties using a suitable measure (or by directly comparing the
different probabilities involved). The output will be a
probabilistic assertion of the form ``$x$ is assigned to the class
$C_{i}$ with probability $p(C_{i})$" (in other words, the output
will be a vector $(p(C_{1}),...,p(C_{m}))$, with complete certainty
when a particular $p(C_{i})$ is one and the rest is zero). When
probabilities are involved, we will refer to it as the
\emph{probabilistic} classical pattern recognition problem. If there
are no probabilities involved, we will say that the problem is
\emph{deterministic}.

\section{Non-Kolmogorovian Measures And Convex Operational Models}\label{s:Generalized Probabilities}

In the above formulation of the pattern recognition problem, it is
assumed that the properties involved are classical. In other words,
the objects are assumed to possess a collection of properties which
can assume definite values at the same time. But if the objects
involved obey the laws of quantum mechanics, then, incompatible
properties may come into play. As it is well known, the
Kochen-Specker theorem precludes the possibility of assigning states
defining simultaneous definite properties to quantum systems (see
for example \cite{Kochen-Specker} for standard quantum mechanics and
\cite{Doring-KSVNA} for general von Neumann algebras). Consequently,
the probabilistic measures involved will no longer be Kolmogorovian
\cite{Redei-Summers2006}: probabilities are now assigned to
elements in the orthomodular lattice of projection operators in a
Hilbert space \cite{vN}. In the following we review the formal
structure of these non-classical features.

A suitable framework to begin the study of non-Kolmogorovian
probabilities is that of measures in \emph{orthomodular lattices}
\cite{Adilee-CopulasQL,belcas81,Hamhalter-QuantumMeasureTheory,kalm83,Ledda-Sergioli-2010}
(see also \cite{Aerts-Introducing}, for a formulation of
non-Kolmogorovian probabilities as functions valued in subsets
instead of numbers). Suppose that an algebra of events can be
represented as an orthomodular lattice $\mathcal{L}$.\footnote{An
\textit{orthomodular lattice} $\mathcal{L}$, is an orthocomplemented
lattice satisfying that for any $a$, $b$ and $c$, if $a\leq c$, then
$a\vee(a^{\bot}\wedge c)=c$. We refer the reader to \cite{kalm83}
for a detailed exposition.} Then, a generalized probabilistic
measure can be represented as a function
\cite{dallachiaragiuntinilibro}:

\begin{eqnarray}\label{e:GeneralizedProbability}
&\nu:\mathcal{L} \rightarrow [0,1],&\nonumber\\
&\mbox{such that}&\nonumber\\
&\nu(\mathbf{1})=1,&\nonumber\\
&\mbox{and, for a denumerable and pairwise orthogonal family of events}&\nonumber\\
&\{E_{i}\}_{i\in I},&\nonumber\\
&\nu(\sum_{i\in I}E_{i}) = \sum_{i\in I}\nu(E_{i}) \,.&\nonumber\\
\end{eqnarray}

\noindent When $\mathcal{L}$ is a Boolean algebra, the above axioms
reduce to the well known Kolmogorov's axioms for classical
probability calculus. In this framework, the elements of a Boolean
algebra are intended to represent properties of a classical system.
On the other hand, if $\mathcal{L}$ represents the orthomodular
lattice of projection operators acting on the Hilbert space of a
quantum system,\footnote{In the Hilbert space case, projection
operators are in one to one correspondence to closed subspaces
(thus, these notions are interchangeable). Representing ``$\vee$" by
the closure of the sum of two subspaces, ``$\wedge$" by its
intersection, ``$(...)^{\bot}$" by taking the orthogonal complement
of a given subspace and ``$\leq$" by subspace inclusion, it is
possible to show that subspaces (and thus, projections) possess an
orthomodular lattice structure.} we recover
--- via the celebrated Gleason's theorem
\cite{Gleason-Dvurechenski-2009,Gleason} --- the probability
assignment given by Born's rule: if a quantum system is prepared in
state $\rho$, the probability of observing the property represented
by projection $P$ is given by

\begin{equation}
p_{\rho}(P)=\mbox{tr}(\rho P).
\end{equation}

\noindent Typically, more general theories of interest can be
considered, as for example, the projection lattices of algebraic
relativistic quantum field theory, which involves projection
lattices of Type III factors \cite{Redei-Summers2006}, or algebraic
quantum statistical mechanics \cite{Redei-Summers2006}. We want to
stress that the underlying algebraic structure determines the
structural features of the probabilities and correlations involved,
and this defines essentially different pattern recognition models.

We will discuss some examples of this in the following sections. As
is well known, observables in quantum mechanics can be represented
as operators in a Hilbert space, being a non-commutative algebra
their most distinctive feature \cite{vN}. This is strongly related
with the non-Booleanity of the lattice of projection operators
\cite{belcas81,vN}. In the next section, we present a version of the
pattern recognition problem for non-Kolmogorovian probabilistic
measures which includes the quantum and classical cases as
particular instances.

It is easy to show that the set of states defined by Eqns.
\ref{e:GeneralizedProbability} is convex. This feature can be taken
as the starting point for a more general approach to the study
statistical theories, based on the geometrical properties of convex
sets \cite{Barnum-Wilce-2006,Barnum-Wilce-2010,Barnum-Wilce-2009}.
Assume that the set of states of a given model is represented by a
convex set $\mathcal{S}$. Then, for each observable with outcome set
$X$, a given state $s\in\mathcal{S}$ should define a probability
$p(s,x)$ for each possible outcome $x\in X$. Given a state
$s\in\mathcal{S}$ and any outcome $x\in X$, it is natural to define
an affine evaluation-functional $f_{x}:\mathcal{S}\rightarrow [0,1]$
by $f_{x}(s):=s(x)$ (where $s(x)$ is a real number in the interval $[0,1]$ that represents the evaluation). Then, it is reasonable to consider each
functional $f:\mathcal{S}\rightarrow [0,1]$ as representing a
measurement outcome, and thus represent that outcome by $f(s)$ (if
the state of the system is $s$). In this way, states are interpreted
as points of a convex set, embedded in a vector space
$V(\mathcal{}S)$, and observables (called \emph{effects} in the
generalized setting) as continuous linear functionals in the dual
space $V^{\ast}(\mathcal{S})$ acting on this set. It turns out that
the shape of the convex set has information about the model
involved. For example, the faces of the convex set of a quantum
system define an orthomodular lattice which is isomorphic to the
lattice of projection operators, while for classical systems, the
set of faces forms a Boolean lattice \cite{Bengtsson2006}. In this
way, for some important models, it is possible to relate the
approach based on measures over lattices with the approach based on
convex sets. Notions like those of pure and mixed states,
entanglement and information, are defined in a natural way, which
generalizes the quantum scenario. We refer the reader to
\cite{Barnum-Wilce-2006,Barnum-Wilce-2010,Barnum-Wilce-2009} for
more details.

\section{Pattern Recognition In The Generalized
Setting}\label{s:QuantumPatternRecognition}

Let us now introduce our general framework for dealing with the
quantum pattern recognition problem in generalized probabilistic
models. It will contain the quantum and classical versions of the
problem as particular cases.

Given a collection of classes of objects $O_{i}$, let us assume that
the state of each object $o^{i}_{j}$ (i.e., object $j$ of class
$O_{i}$) is represented by a state $\nu^{i}_{j}\in \mathcal{C}_{i}$,
where $\mathcal{C}_{i}$ is the convex operational model representing
object $o^{i}_{j}$. We will assume that all objects in the class
$O_{i}$ are represented by the same convex operational model
$\mathcal{C}_{i}$ (i.e., they are all elements of the same type).
Then, suppose that weights $p^{i}_{j}$ are assigned to the objects
$o^{i}_{j}$, representing the rational agent's knowledge about the
importance of object $o^{i}_{j}$ as a representative of class
$\mathcal{C}_{i}$ with respect to other objects in the class (if all objects are equally important, the weights
are chosen as $p^{i}_{j}=\frac{1}{N_{i}}$, where $N_i$ is the total number of objects - i.e. the cardinality - of the class $\mathcal{C}_{i}$). This means that the
probabilistic state of the whole class $O_{i}$ can be represented by
a mixture $\nu_{i}=\sum_{j}p_{j}\nu^{i}_{j}\in \mathcal{C}_{i}$. As
we discuss below, it is also possible to assume that non-local
correlations are given between the different classes, and the states
$\nu_{i}$ are reduced states of a global --- possibly entangled ---
state $\tilde{\nu}$. But we notice that under these conditions, the
states $\nu_{i}$ will be \emph{improper mixtures}, and then, no
consistent ignorance interpretation can be given for them
\cite{d'esp}.

The generalized pattern recognition problem is then posed as
follows. A particular object $o$ must be identified and compared
with the information given by the generalized states of the classes
represented by $\nu_{i}$ (or more generally, by $\tilde{\nu}$),
obtained in the learning process. The comparison could be also
restricted to a collection of properties
$\vec{a}=(\alpha_{1},....,\alpha_{m})$, represented now by
generalized effects $\alpha_{i}$. We will assume, as usual, that
knowledge about $o$ is represented by a generalized state $\nu$.
Notice that, in order to obtain $\nu$, several copies of the unknown
object $o$ may be needed, whenever the probabilistic character of
the model is irreducible. This is the case in quantum mechanics: if
more copies are available, the reconstruction of the state of the
unknown object will be more accurate, and this can be used to
improve the classification process.

Different techniques for discriminating the given state with regard
to the states of the classes were studied for some particular models
(see for example \cite{QPR3-Guta-Wojciech-NJP} and
\cite{Sentis-CoherentStates-2015}). Notice that optimal
classification strategies may depend strongly on the structural
properties of the probabilities associated to the model involved.
Notice also that non-classical correlations between the classes
represented by states $\nu_{i}$ may come into play (the $\nu_{i}$
may be reduced states of a global state $\mu$). Again, the
particularities of the correlations originated in each model may be
critical here (see for example
\cite{Clifton-Halvorson-EntanglementARQFT}, for a discussion of the
differences between relativistic and non-relativistic quantum
entanglement).

It is easy to see that, if the $\mathcal{C}_{i}$'s are simplices
(hyper-tetrahedrons), then, we will recover the classical problem of
patterns' recognition, discussed in the introduction. Indeed,
probabilities on simplices are isomorphic to measures over Boolean
algebras, and thus, we have Kolmogorovian probabilities. And this is
nothing but our definition of classical pattern recognition problem.
As it is well known, simplices admit dispersion-free states: this
means that using this description, we can also recover the
deterministic version of the problem described in Section
\ref{s:IntroPatternRcognition}. On the other hand, as we remarked in
Section \ref{s:Generalized Probabilities}, the sets of states of
quantum systems are naturally convex sets, but are not simplices
\cite{Bengtsson2006}.

\section{Examples}\label{s:Examples}
\noindent

As examples of the general framework introduced above, in this
Section we briefly describe non-classical examples of the pattern
recognition problem which originate in non-equivalent physical
theories of interest.

\subsection{Quantum Pattern Recognition}

Suppose that we are given a collection of quantum objects each
belonging to a particular class $Q_{i}$, and given a particular
object $q$, the rational agent aims to determine to which class it
is assigned. We look now for a quantal version of the problem posed
in Section \ref{s:IntroPatternRcognition}. First, we must assume, as
the most general possibility, that the collection of chosen
properties can be non-commutative. Thus, the properties of object
$q^{i}_{j}$ (object $j$ of class $C_{i}$) will be represented by
operators\footnote{Notice that these operators could be quantum
effects without loss of generality.} acting on a Hilbert space
$\mathcal{H}_{i}$ (representing the class $Q_{i}$). It is now
impossible (in the general case) to assign a vector of definite
properties to each object, due to possibly non-commutativity of the
operators involved. The only thing that we can do, is to assign
probabilities for each property coordinate using the quantum state
$\rho^{i}_{j}$ of each object $q^{i}_{j}$. Thus, if --- as in the
classical case --- we assign weights $p^{i}_{j}$ to each object
$q^{i}_{j}$, knowledge about the class $Q_{i}$ can now be
represented by a mixture $\rho_{i}=\sum_{j}p^{i}_{j}\rho^{i}_{j}$.
Given the fact that in general, interaction between physical systems
represented by classes $Q_{i}$ can be non-negligible, and thus,
non-trivial correlations may be involved, we will assume that the
states $\rho_{i}$ are arbitrary states of the Hilbert space
$\mathcal{H}_{i}$ (i.e., the $\rho_{i}$ are not necessarily proper
mixtures). We call $\tilde{\rho}$ the global state of the whole set
of classes.

Given an arbitrary individual $q$, we are thus faced with the
problem of determining to which class $Q_{i}$ it should be assigned.
In the general case, the state of $q$ will be represented by a
density operator $\rho$ (acting on one of the unknown Hilbert spaces
$\mathcal{H}_{i}$, but certainly embedded in the Hilbert space
$\mathcal{H}_{1}\otimes\mathcal{H}_{2}...\otimes\mathcal{H}_{n}$).
Notice however, that the state $\rho$ could, in the general case, be
unknown to the agent, and he may have only access to a sample of
values $\{a_{j}\}$ of the operators $\sigma_{j}$. Thus, for the
classification problem, he should be able to, either reconstruct the
unknown state $\rho$ using quantum statistical inference methods, or
just directly compare the sampled values with the information
provided by the global state $\tilde{\rho}$.

Our version of the quantum pattern recognition problem can be
interpreted in a very direct physical way. What we have shown in
this section is that if the objects involved exhibit non-classical
features (and this could be the case each time that the systems
involved are small enough to exhibit quantum behavior), then, the
rational agent will be confronted with complementarity phenomena,
non-Kolmogorovian probability measures, and non-classical
correlations. In this way, the information about the parties
involved must be necessarily represented by density operators.
Furthermore, as we have seen, the classification problem must be
adapted to this situation in such a way that quantum statistical
inference techniques must be used in order to decide which will be
the class to which the object will be most likely assigned.

From this physical perspective, how is it possible to represent
information updating and learning? In other words, which is the
quantum analogous of a semi-supervised system? This can be suitably
described using quantum operations as follows. Suppose now that at
an initial state, the agent has an information $\rho_{i}(0)$ for
each class $Q_{i}$, and he is confronted with an individual of which
it has information $\rho(0)$, and a global state $\tilde{\rho}(0)$.
Then, after the classification process at time $t$, it is necessary
to update knowledge about the classes and the global state to new
states $\rho_{i}(t)$ and $\tilde{\rho}(t)$,  respectively. This can
be suitably modeled by a quantum operation $\Lambda(t)$ acting on
the convex quantum set of states of
$\mathcal{C}(\mathcal{H}_{1}\otimes\mathcal{H}_{2}...\otimes\mathcal{H}_{n})$,
such that $\Lambda(t)\tilde{\rho}(0)=\tilde{\rho}(t)$. A
\emph{quantum learning operator} will be thus a family of quantum
operations $\{\Lambda(t_{1}),...,\Lambda(t_{n})\}$. Hence, a
\emph{quantum learning process} will be a succession of global
states
$\{\tilde{\rho}(0),\Lambda(t_{1})\tilde{\rho}(0),\Lambda(t_{2})\tilde{\rho}(t_{1}),...,\Lambda(t_{n})\tilde{\rho}(t_{n-1})\}$.
The goal of the learning process will be achieved if the uncertainty
of the final state is reduced. The dispersion could be measured
using the von Neumann entropy (or other quantum entropic measures)
\cite{Holik-Bosyk-QINP-2016,HolikQIC-2016,Holik-Entropy-2015}.

This dynamical view of the quantum learning process can be easily
generalized to arbitrary statistical models by appealing to affine
maps. The entropies used to measure the success of the learning
process could be the Measurement Entropy (or any other entropic
measure of interest which can be suitably generalized
\cite{HolikQIC-2016,Holik-Entropy-2015}).

Let us also notice that in the case of unsupervised learning - where the training data are not labeled - the classes of the classification process are, in principle, unknown and the learning process is fundamental in order to individuate the distribution of these classes. If the assignment of each object to its respective class is given by a Kolmogorovian probability function, then the distribution of the classes that arises from the learning process is a kind of partition such that each member of the dataset is classified within one and only one class; the Voronoi diagram is an instance of this kind of partition. On the other hand, if the probabilities involved to describe the objects in question are non-Kolmogorovian, the distribution of the classes that arises from the learning process has to take into account this change in the mathematical description. For the quantum case, the mathematical description of the states of the objects in the dataset is given in terms of vectors in a Hilbert space (or more generally, density operators), while their properties are represented by self-adjoint operators (see \cite {Aimeur-QuantumWorld}).

\subsection{Pattern Recognition In ARQFT}

In algebraic relativistic quantum field theory (ARQFT), a
$C^{\ast}$-algebra is assigned to any open set $O$ of a differential
manifold $M$ \cite{HalvorsonARQFT,HaggLQP}. Open sets are intended
to represent local regions, and $M$ models space-time with its
symmetries. Local algebras are intended to represent local
observables (such as particle detectors). For example, in ARQFT, $M$
is Minkowski's four dimensional space-time, endowed with the
Poincare group of transformations.

It turns out, that (global) states of the field define measures over
the local algebras. But in general, the local algebras of ARQFT will
not be Type I factors as in standard quantum mechanics. For example,
it can be shown that for a diamond region, a Type III factor must be
assigned \cite{Yngvason-TypeIII}. This means that the orthomodular
lattice involved in axioms \ref{e:GeneralizedProbability}, will not
be the lattice of projection operators of a Hilbert space, but a one
with different properties. For a discussion on the properties of
lattices associated to von Neumann algebras see
\cite{mikloredeilibro}, Section $6.2$. Consequently, operator
algebras in ARQFT are quite different to those of standard quantum
mechanics. This is expressed, for example, in the properties of
correlations (see for example
\cite{Clifton-Halvorson-EntanglementARQFT} and Chapter $3$ in
\cite{HalvorsonARQFT}).

This means that the discrimination problem must be posed between
classes $F_{i}$ represented by states of the field $\varphi_{i}$ and
a given individual state $\varphi$.\footnote{In practical
implementations, these states and the discrimination problem, could
be restricted to a concrete space-time region.} As far as we know,
this problem was not addressed in the literature from the point of
view of pattern recognition. But it is an important one, because in
the general case, it could be useful for information protocols based
on quantum optics (where the effects of the field character of the
theory cannot be neglected). In particular, a simpler but analogous
version of the problem could be conceived by appealing to the
Fock-space formalism, in order to describe the fields and the states
involved (see for example \cite{Sentis-CoherentStates-2015}, for a
version of the problem posed in terms of coherent states of light
using the Fock-space formalism).

\subsection{Pattern Recognition In AQSM}

As in the quantum field theoretic example, a similar problem can be
posed in the algebraic approach to quantum statistics (AQSM). Here,
a typical problem could be to discern a kind of atoms from a set of
classes of gasses; now, the comparison will be between the state of
the item and the classes involved. But it can be shown that the
global states of a gas, as described by AQSM, will be in general, a
measure over a factor different from the Type I case (see for
example \cite{bratelli}, Chapter $5$ and \cite{Redei-Summers2006}).
This means that, again, the pattern recognition problem will depart
from that of the classical one, but also from that of standard
quantum mechanics (where we have Type I projection lattices).

An example of interest could appear in problems related to image
recognition. To clarify ideas, let discuss first a classical version
of the problem. Suppose that a machine has to solve a problem of
recognizing handwritten digits. These drawings are first transformed
into digitalized images of $n\times n$ pixels. This means that the
information of each image is stored in a vector $\vec{x}$ of length
$n\times n$. The goal is to build our automata in such a way that it
takes a vector $\vec{x}$ as an input, and gives us as output the
identity of the digit in question \cite{Bishop-Book}. Now we pose
the question: in a real hardware, this vector should be stored using
bits of a given length. But if the components of the hardware are so
small that they become quantal (as in \cite{Guang-2016}), then, we
may have a large chain of qubits used to store this information.
This means that the information will be stored now in a potentially
large chain of qubits $\bigotimes^{n}_{i=1}\mathbb{C}^2_{i}$. When
$n$ is big enough, the number of particles involved to store the
information may become statistical. Thus, the approximation of
$n\longrightarrow\infty$ becomes more and more realistic in order to
represent global properties of the information stored. But in this
limit, we need the algebras corresponding to the algebraic
formulation of quantum statistical mechanics.

Let us illustrate this with a concrete model. Suppose that we have a
spacial arrangement $L$ of $N$-dimensional quantum systems. For each
point $x\in L$ we have a Hilbert space $\mathcal{H}_{x}$, and for
each subset of points $\Gamma\in L$, the associated Hilbert space is
given by the tensor product
$\mathcal{H}_{\Gamma}=\bigotimes_{x\in\Gamma}\mathcal{H}_{x}$. Thus,
every subset $\Gamma\in L$ has associated an algebra
$\mathcal{A}(\mathcal{H}_{\Gamma})$. The norm completion of the
collection $\mathcal{A}=\{\mathcal{A}_{\Gamma}\}_{\Gamma\in L}$ is a
quasi-local $C^{\star}$-algebra when equipped with the net of
$C^{\star}$-subalgebras $\mathcal{A}_{\Gamma}$. Thus, the
classification problem must be done with respect to states defined
in this algebra (such as KMS-states \cite{bratelli}), whose
properties are different to that of a Type I factor.

\section{Quantum Algorithms As Quantum Pattern Recognition Problems}\label{s:Computational complexity}

\noindent

Recent developments suggest that quantum speedups appear in
structured problems \cite{Aaronson-2014}: the problem must exhibit
some structure or pattern in order that the quantum computer display
an overhead with respect to a classical one. Indeed, in
\cite{QPR3-Sasaki-Carlini} the authors suggest that in a certain
sense, the most important quantum computation algorithms can be
viewed as pattern recognition problems. Let us now outline how our
generalized formalism could be useful to formulate these deep
intuitions on a more solid ground, by looking at some examples
of quantum algorithms.

\subsection{Deutsch-Jozsa algorithm}

Let us examine first the Deutsch-Jozsa algorithm \cite{NielsenBook}.
In this case, the task is to determine if a function $f$ is constant
or balanced. There are four functions from
$\{0,1\}\longrightarrow\{0,1\}$, namely:

\begin{eqnarray}
f_{1}(0)=0\,\,\,f_{1}(1)=1\nonumber\\
f_{2}(0)=1\,\,\,f_{2}(1)=0\nonumber\\
f_{3}(0)=0\,\,\,f_{3}(1)=0\nonumber\\
f_{4}(0)=1\,\,\,f_{4}(1)=1\nonumber\\
\end{eqnarray}

\noindent Thus, we have two classes: $C=\{f_{1},f_{2}\}$ and
$B=\{f_{3},f_{4}\}$, and we must determine if the function $f$
belongs to $B$ or to $C$. Up to now, this is just a classical
pattern recognition problem.

Let us see now how the quantum computer transforms this problem into
a quantum pattern recognition one. The computer is prepared first in
the quantum state $|0\rangle|1\rangle$. Next, the Hadamard operator
is applied to both qubits yielding the state:

$$\frac{1}{2}(|0\rangle+|1\rangle)(|0\rangle-|1\rangle).$$

Next, the quantum implementation of the function $f$ (which brings
the connection between the classical problem and the quantum
computation) will be given by a quantum operator such that it maps
$|x\rangle|y\rangle$ to $|x\rangle|f(x)\oplus y\rangle$. Applying
this function to the state gives

$$(-1)^{f(0)}\frac{1}{2}(|0\rangle+(-1)^{f(0)\oplus f(1)}|1\rangle)(|0\rangle-|1\rangle).$$

\noindent Now, applying the Hadamard transformation again to the
first qubit we get:

$$|\psi\rangle=(-1)^{f(0)}\frac{1}{2}((1+(-1)^{f(0)\oplus f(1)})|0\rangle +(1-(-1)^{f(0)\oplus f(1)})         |1\rangle)(|0\rangle-|1\rangle).$$

\noindent The next step consists in determining the projection of
the above state to the subspaces represented by projection operators
$|0\rangle\langle 0|\otimes\textbf{1}$ and $|1\rangle\langle
1|\otimes\textbf{1}$. This is nothing but determining if the system
represented by state $|\psi\rangle\langle\psi|$ belongs to the
classes represented by projections $|0\rangle\langle
0|\otimes\textbf{1}$ and $|1\rangle\langle 1|\otimes\textbf{1}$: the
computation of the projections is nothing but the Hilbert Schmidt
distance between these operators. Thus, this simple problem shows
that this is a pattern recognition problem in which the rational
agent has to decide if an individual (the output state of the
computer previous to measurement) represented by state
$\rho=|\psi\rangle|\psi\rangle$ belongs to the class represented by
state $\rho_{1}=|0\rangle\langle 0|\otimes\textbf{1}$ and the class
represented by state $|1\rangle\langle 1|\otimes\textbf{1}$.

\subsection{Period of a function determination}

The determination of the period of a periodic function $f$ lies at
the heart of the Shor and Simon quantum computation algorithms
\cite{NielsenBook}. Here we show that this problem can be reduced to
a quantum pattern recognition one.

The objective now is to determine the period of a function
$f:\mathcal{Z}_{N}\longrightarrow \mathcal{Z}$, such that
$f(x+r)=f(x)$ for all $x$. It is assumed that the function does not
take the same value twice in the same period. Start the computer as
usual by generating the state:

\begin{equation}
|f\rangle=\frac{1}{\sqrt{N}}\sum_{x=0}^{N-1}|x\rangle|f(x)\rangle.
\end{equation}

It is not possible to extract the period yet. Even if we measure the
value of the second register and obtain the value $y_{0}$, we will
end up with the following state in the first register (with $x_{0}$
the smallest $x$ such that $f(x)=y_{0}$ and $N=K r$):

\begin{equation}
|\psi\rangle=\frac{1}{\sqrt{K}}\sum^{K-1}_{k=0}|x_{0}+k r\rangle .
\end{equation}

But $|\psi\rangle$ does not give us information about $r$ yet. To do
that, it is necessary to apply the quantum Fourier transform (QFT),
which is a unitary matrix with entries

\begin{equation}
\mathcal{F}_{ab}=\frac{1}{\sqrt{N}}\exp^{2\pi i \frac{ab}{N}}.
\end{equation}

\noindent By applying the QFT to $|\psi\rangle$ we obtain

\begin{equation}
\mathcal{F}|\psi\rangle=\frac{1}{\sqrt{r}}\sum^{r-1}_{j=0}\exp^{2\pi
i \frac{x_{0}j}{r}}|j\frac{N}{r}\rangle .
\end{equation}

Now, a measurement is performed in the basis
$\{|j\frac{N}{r}\rangle\}$, and using the result it is possible to
determine the period of the function as follows. The obtained value
$c$ will be such that $c=j\frac{N}{r}$, for some $0\leq j\leq r-1$.
Then, $\frac{c}{j}=\frac{N}{r}$, and if $j$ is coprime with $r$, it
will be possible to determine $r$. The success of the algorithm
depends on the fact that $j$ and $r$ will be coprimes with a high
enough probability.

The key observation here, is that this can be cast as a pattern
recognition problem as stated below. The objective is to decide to
which class pertains an individual (again, the output of the
computer after the second register measurement and application of
the quantum Fourier transform) represented by state
$\mathcal{F}|\psi\rangle\langle \psi|\mathcal{F}$ (with
$|\psi\rangle=\frac{1}{\sqrt{K}}\sum^{K-1}_{k=0}|x_{0}+k r\rangle$).
The classes are represented by states $\{\rho_{i}\}$, with
$\rho_{i}=|j\frac{N}{r}\rangle \langle j\frac{N}{r}|$. From the
expression
$\mbox{tr}(\mathcal{F}|\psi\rangle\langle\psi|\mathcal{F}|j_{0}\frac{N}{r}\rangle
\langle j_{0}\frac{N}{r}|)$, which is the same as
$\mbox{tr}(|\psi\rangle\langle\psi|\mathcal{F}|j_{0}\frac{N}{r}\rangle
\langle j_{0}\frac{N}{r}|\mathcal{F})$, we obtain an equivalent
problem by comparing state $|\psi\rangle\langle\psi|$ with states
$\mathcal{F}|j_{0}\frac{N}{r}\rangle\langle
j_{0}\frac{N}{r}|\mathcal{F}$. We are of course interested in
identifying those measurements for which $\frac{c_{0}}{N}$ is an
irreducible fraction.

\section{Conclusions}\label{s:Conclusions}
\noindent

In this paper we have presented a generalization of the pattern
recognition problem to the non-commutative (or equivalently,
non-Kolmogorovian) setting involving incompatible
(non-simultaneously determinable) properties. In other words, we
have cast the problem of pattern recognition for the case in which
the state spaces involved are not simplexes. In this way, we have
shown that it is possible to find some important (and
non-equivalent) examples of interest: standard quantum mechanics,
algebraic relativistic quantum field theory, and algebraic quantum
statistics. The examples do not restrict only to these ones, but can
include more general models, and particular, hybrid systems
(classical and quantum). In particular, studies such as
\cite{Abramsky}, suggest that the study of the pattern recognition
problem in non-Kolmogorovian probabilistic models could, in
principle, turn out to be particularly beneficial for the treatment
of relational databases.

Next, we have shown hat our perspective could be useful to
characterize some of the most important quantum computation
algorithms (Shor, Simon and Deutsch-Jozsa) as quantum pattern
recognition problems. This may also suggest that it is to be
expected that, translating classical pattern recognition problems
into quantum ones, could lead to an improvement in the efficiency of
the concomitant computation.

Our framework allows for clear definitions of the classical and
quantum pattern recognition problems, respectively, and for an
extension of the applicability of the problem to a wider domain.

\bigskip

\textbf{Acknowledgments}

This work was partially supported by CONICET and UNLP (Argentina),
the project ``Computational quantum structures at the service of
pattern recognition: modeling uncertainty" $[CRP-59872]$ funded by
Regione Autonoma della Sardegna, L.R. $7/2007$, Bando $2012$ and the
FIRB project Structures and Dynamics of Knowledge and Cognition,
Cagliari: $F21J12000140001$, founded by  Italian Ministery of
Education. The authors thank anonymous reviewers for useful
comments.

\end{document}